\documentclass[10pt,aas,apj,superscriptaddress]{emulateapj}

\usepackage{amsmath}
\usepackage{epsfig}
\usepackage{enumerate}
\usepackage{graphicx}
\usepackage{hyperref}
\usepackage{url}
\usepackage{natbib}
\newcommand{\msun}{{\rm M}_{\odot}}	
\newcommand{\mbh}{M_{\bullet}}		
\newcommand{\nbh}{n_{\bullet}}		
\newcommand{\ms}{M_{\star}}		
\newcommand{\mbhc}{\mathcal{M}_{\bullet}}	
\newcommand{\nbhc}{\varphi_{\bullet}}	
\newcommand{\msc}{\mathcal{M}_{\star}}		
\newcommand{\nsc}{\varphi_{\star}}		
\newcommand{\mc}{\mathcal{M}}			
\newcommand{\nc}{\varphi}		
\newcommand{\rbh}{\rho_{\bullet}}	
\newcommand{\rs}{\rho_{\star}}		
\newcommand{\mlow}{10^6\,\msun}
\newcommand{\mhi}{10^{11}\,\msun}
\newcommand{\al}{\alpha} 
\newcommand{\albh}{\alpha_{\bullet}} 
\newcommand{\als}{\alpha_{\star}}
\newcommand{\dint}{\displaystyle\int}

\newcommand{\ie}{i.\,e. } 
\newcommand{\eg}{e.\,g. }
\newcommand{\beq}{\begin{equation}}
\newcommand{\eeq}{\end{equation}}
\newcommand{\bea}{\begin{eqnarray}}
\newcommand{\eea}{\end{eqnarray}}

\def\leq{\,\raise 0.4ex\hbox{$<$}\kern -0.8em\lower 0.62ex\hbox{$-$}\,}
\def\geq{\,\raise 0.4ex\hbox{$>$}\kern -0.7em\lower 0.62ex\hbox{$-$}\,}
\def\lsim{\,\raise 0.4ex\hbox{$<$}\kern -0.8em\lower 0.62ex\hbox{$\sim$}\,}
\def\gsim{\,\raise 0.4ex\hbox{$>$}\kern -0.7em\lower 0.62ex\hbox{$\sim$}\,}
\def\apppropto{\,\raise 0.4ex\hbox{$\propto$}\kern -0.7em\lower 0.62ex\hbox{$\sim$}\,}

\begin{document}

\title{Gravitational waves and stalled satellites from massive galaxy mergers at $z\leq1$}
\author{Sean T. McWilliams}
\affiliation{Department of Physics, Princeton University, Princeton, NJ 08544}
\email{stmcwill@princeton.edu}
\author{Jeremiah P. Ostriker}
\affiliation{Department of Astrophysical Sciences, Princeton University, Princeton, NJ 08544}
\author{Frans Pretorius}
\affiliation{Department of Physics, Princeton University, Princeton, NJ 08544}



\begin{abstract}
We present a model for merger-driven evolution of the mass function for massive galaxies and their central supermassive black holes at late times.  We
discuss the current observational evidence in favor of merger-driven massive galaxy evolution during this epoch, and
demonstrate that the observed evolution of the mass function can be reproduced by evolving an initial mass function
under the assumption of negligible star formation.  We calculate the stochastic gravitational wave signal 
from the resulting black-hole binary mergers in the low redshift universe ($z\leq 1$) implied by this model,
and find that this population has a signal-to-noise ratio as much as $\sim 5\times$ larger than previous estimates for pulsar
timing arrays, with an expectation value for the characteristic strain $h_{\rm c}(f=1\,{\rm yr}^{-1})=4.1\times 10^{-15}$ 
that may already be in tension with observational constraints, and a \{2$\sigma$, 3$\sigma$\} lower limit within 
this model of $h_{\rm c}(f=1\,{\rm yr}^{-1})=$ \{$1.1\times 10^{-15}$, $6.8\times 10^{-16}$\}.
The strength of this signal is sufficient to make it detectable with high probability
under conservative assumptions within the next several years, if the principle assumption of merger-driven galaxy evolution since
$z=1$ holds true.  For cases where a galaxy merger fails to lead to
a black hole merger, we estimate the probability
for a given number of satellite unmerged black holes to remain within a massive host galaxy, and interpret the result in light of ULX observations.
In particular, we find that the brightest cluster galaxies should have 1--2 such sources with luminosities above
$10^{39}$ erg/s, which is consistent with the statistics of observed ULXs.
\end{abstract}

\maketitle

\section{Introduction}
\label{sec:intro}
The growth of supermassive black holes (SMBH) at the centers of massive galaxies is thought to have occurred primarily at large redshifts,
with the cessation of quasar activity at $z\leq 2$ signaling the end of SMBH growth as well (see \cite{Natarajan} for a review).  
More specifically, SMBHs are thought to grow primarily through gas accretion at higher redshifts; gas-rich galaxy mergers
would trigger the accretion episodes through which SMBHs would grow, with the actual merging of multiple SMBHs being a subdominant
mode of SMBH growth \citep{VolNat,Treister}.  At lower redshifts, gas depletion would result in a greater relative importance
of merging black holes to SMBH mass growth.  This epoch is associated with lower rates of growth in SMBH mass density overall.
However, mergers can redistribute the mass density at late times ($z\leq 1$), leading to significant growth of the most massive black holes,
even in the absence of observable amounts of gas accretion.  The increased importance of mergers at late times, if true, will have two
inescapable consequences: the stochastic gravitational-wave background from merging SMBHs will be larger than otherwise expected,
and any black holes that fail to merge will spiral through their host galaxy indefinitely, accreting gas and stars and emitting
X-rays as they go.

Recent evidence indicates that, although these mostly ``red and dead'' massive galaxies undergo little star formation at late times,
they continue to merge unabated, so that mergers dominate the evolution of the black-hole mass function for large masses at $z\leq 1$.  
Although some papers have indicated that massive black holes grow by as little as $\sim 1/3$ of their original mass since $z=1$ 
(e.~g.~\cite{Brown,Yoo,deRavel}), more recent results of both observational campaigns \citep{Zhang,Laporte} and numerical simulations \citep{Oser2012} 
strongly suggest that massive galaxies have grown still more rapidly, and may have as much as doubled their mass since $z=1$.  
We will adopt the assumption of mass doubling since $z=1$, with the obvious caveat that a smaller amount of 
growth for massive ellipticals will imply a proportionally smaller number of merger events.  However, since the gravitational
wave signals from these events will add incoherently, we expect the full range of predicted merger rates to cause no more
than a $\sim \sqrt{1/3}$ difference in the predicted signal.
Both observations and simulations further indicate that the growth of massive galaxies since $z=1$ occurred
primarily through mergers with $M^s_{\bullet}/M^h_{\bullet}\geq 1/10$, where the black hole of mass $M^s_{\bullet}$ 
from the satellite galaxy merges with the host black hole of mass $M^h_{\bullet}$
through dynamical friction, stellar hardening (or some alternative process), 
then gravitational wave emission, and the gas and stars from the satellite
are deposited in the outer reaches of the host galaxy.  

The anomalously large optical luminosity observed in the brightest cluster galaxies (BCGs) has long suggested that mergers play
a critical role in their evolution, providing another indication of the importance of mergers for massive galaxies.
The Bautz-Morgan classification \citep{Bautz:1970} of galaxy clusters depends primarily on the magnitude difference, $\Delta m_{1\,2}$, between the first
and second brightest galaxies in a cluster; the large size of this difference in some clusters
had been a noted anomaly (see \cite{Abell:1965} and references therein).
\cite{TremaineRichstone} invented a statistical test that implicated a physical process, rather than a statistical variation,
as the cause of the large value of $\Delta m_{1\,2}$.  \cite{Ostriker:1977} and \cite{Hausman:1978} argued that a plausible origin
was ``galactic cannibalism'' driven by the process of dynamical friction, which would cause a massive satellite galaxy to spiral into the
center of a cluster and merge with the BCG residing there.  Many subsequent studies have affirmed the likelihood of this scenario.

In this work, we will discuss the observable consequences of the fact that these merging massive galaxies will generally contain
SMBHs.  The satellite black holes will be detectable as accreting X-ray sources as they orbit within the cannibal BCG, and then
as gravitational wave sources if they reach the more massive black hole at the center of the BCG.
We will demonstrate that the observed evolution of the galaxy and black-hole mass functions can be reproduced by a calculation
that assumes only mergers drive the evolution.  By comparing the calculation with the observed evolution, we can find the total
number of mergers necessary to reproduce the observations.  With this number in hand, we perform a Monte Carlo simulation
of the population of merging black holes, from which we can directly extract the total expected gravitational wave signal.  Our results,
which are the first to be derived primarily from observational data\footnote{After completion of this manuscript, we became aware of
an independent calculation of the gravitational-wave signal from PTAs \citep{Sesananote}, which combines the observed galaxy mass function and pair fraction
to predict the gravitational wave signal.  This new estimate is moderately larger than previous estimates.},
indicate that the signal within the frequency band of pulsar timing arrays (PTAs) is substantially larger than previously estimated,
and is in fact at the threshold of detectability for the current generation of PTAs.  Furthermore, the signal extends to higher frequencies
than previous estimates, potentially reaching the lowest sensitive frequencies for space-based gravitational wave observatories.
In addition to the galaxy mergers that yield gravitational wave signals, many galaxy mergers will result in what we will call a ``stalled''
satellite, where the smaller black hole cannot make its way to the host core in less than a Hubble time.  We calculate this population
as well, and compare their quantity and expected characteristics to those observed in ultraluminous X-ray sources (ULXs).

In Sec.~\ref{sec:mf}, we will describe our model for the black-hole mass function, first presenting the explicit redshift-dependent
expressions based primarily on observational data in Sec.~\ref{sec:sch}, then discussing our construction of a set of evolution equations
for propagating an initial mass function to later redshifts without explicitly enforcing any redshift dependence in Sec.~\ref{sec:ev}.
We calculate the dynamics of the black holes after galaxy merger to determine whether the black holes will merge in Sec.~\ref{sec:dyn}.
We then apply our model to first calculate the stochastic gravitational-wave signal from merging SMBH binaries in Sec.~\ref{sec:gw}, elaborating
on the results presented in \cite{MOP}.
We then use our model to predict the number of stalled satellite black holes from failed mergers and their potential appearance as ULXs in Sec.~\ref{sec:ulx}.
We summarize and conclude in Sec.~\ref{sec:conc}.
 
\section{The Mass Function}
\label{sec:mf}
\subsection{The Schechter Parameters}
\label{sec:sch}
Any attempt to calculate the stochastic background of gravitational waves from SMBHs requires a model for the
mass and redshift distribution of galaxy mergers.  In order to anchor our model in reality, we wish to calibrate it using
current observations of the mass and redshift distribution of galaxies, typically referred to as the galaxy stellar mass function.  
The information contained in observations is condensed by fitting the data to established models of the mass function, which are
generally empirical.  The stellar bulge and black hole mass functions are frequently parameterized using a Schechter function
of the form \citep{Schechter},
\beq
\phi\,(M)\, dM = \nc \,M^{\al}\, \exp(-M)\, dM\,,
\label{eq:sch}
\eeq
where $\phi\,dM \equiv (\partial n/\partial M) \,dM$ is the comoving number density of either black holes or bulges with masses between
$M$ and $M+dM$, with $M\equiv \mbh/\mbhc$ for black holes or $\ms/\msc$ for bulges. $\mc$, $\nc$, and $\al$ are the three Schechter parameters,
and the subscripts ``$\bullet$'' and ``$\star$'' refer to black holes and stellar bulges, respectively.
However, Eq.~\ref{eq:sch} fails to accurately reproduce the observed mass function when BCGs are included.  In order to
represent both BCGs and less massive galaxies (subscripted ``low''), we combine Eq.~\ref{eq:sch} with a Gaussian component representing
BCGs \citep{Lin}, 
\bea
\phi\,(M)\, dM &\equiv& (\phi_{\rm low}+\phi_{\rm BCG})\,dM = \nc \,M^{\al}\, \exp(-M)\, dM \nonumber \\
&+& \nc \exp\left[-\frac{1}{2}\left(\frac{2.5\log M}{\sigma_M}\right)^2-1\right]\,dM\,,
\label{eq:mf}
\eea
where the single new parameter, $\sigma_M$, determines the e-folding of the high-mass, BCG-populated portion of the mass function.
To avoid ambiguity, we will use $\phi_{\rm tot}\equiv \phi$ to represent the full mass function in any expressions that include both the full function
and the individual components.
We use $\sigma_M \equiv 0.35$ at all redshifts, since there are no apparent physical grounds for $\sigma_M$ to vary, and this value
guarantees the inclusion of ultramassive black holes like M87 within our distribution.  

Since observational data are generally available for stellar bulges rather than for massive black holes directly, we relate
the Schechter parameters for the central black holes
to the parameters for their concomitant stellar bulges by assuming a redshift-independent $\mbh$-$\ms$ relation given by
$\mc \equiv \mbhc = 1.6\times 10^{-3}\msc$ \citep{mbhms,Yu}, $\al \equiv \albh = \als$, and $\nc \equiv \nbhc = \nsc$, which are the only
choices consistent with negligible star formation and negligible accretion of gas onto the central black holes.
We emphasize that recent observational results indicate that simple scaling
relationships like $\mbh$-$\ms$ may systematically underestimate the mass of black holes in BCGs by a factor as large as 10 \citep{Hlavacek}, so we would be equally justified
in enhancing the high-mass portion of the distribution in a similar fashion by including a mass dependence in the proportionality constant of the $\mbh$-$\ms$ relation.
There are a number of other corollaries for merger-driven evolution:
the black-hole mass-stellar-bulge mass ($\mbh$-$\ms$) relation and the black-hole and stellar-bulge densities,
$\rbh$ and $\rs$, are all redshift independent; $\mc(z=0) \geq \mc(z>0)$; and $\al(z=0) \geq \al(z>0)$.

We note that the observational data on the redshift dependence of the Schechter parameters have substantial variance,
which provides the motivation for developing theoretically-motivated dependencies to inform choices for our model.  
For instance, although we assume a constant $\mbh$-$\ms$ relationship,
the literature varies wildly, using $\mbh \approx 10^{-3}\ms (1+z)^a$ with $a$ as large as 2, 
although larger values are typically drawn from fits to data that span to higher redshifts.  
For $\mc$, we use 
\beq
\mc \equiv \frac{1.2\times 10^{8}}{1+z}\,\msun\,.
\eeq
Though consistent with the observational data indicating mass-doubling since $z=1$ \citep{Robaina,vanDokkum}, this 
choice of redshift dependence also makes sense in light of our assumption of merger-driven evolution, since
massive galaxies with $\ms\sim \mathcal{O}(\msc)$ should become more massive at lower redshifts as they consume
less massive satellites.

We choose a constant value for $\nc$,
\beq
\nc \equiv 3 \times 10^{-3}\, {\rm Mpc}^{-3}\,,
\eeq
which is consistent with the literature \citep{Cole2001} at low redshifts when the final Wilkinson Microwave 
Anisotropy Probe (WMAP) cosmological parameters are used \citep{WMAP7}, and is also consistent with merger-driven evolution,
except in the unlikely scenario that most mergers occur between nearly equal mass binaries.
Finally, to find $\al$, we first integrate the mass function for total comoving mass density, 
\beq
\displaystyle\int_0^{\infty} M \nbh\, dM = \mc \nc \Gamma[\al + 2] = \langle\rbh \rangle = {\rm const}\,.
\eeq
In choosing a mean mass density $\langle\rbh \rangle$ for black holes during this epoch, we adopt a conservative estimate \citep{Yu}, 
but correct it to account for the more recent $\mbh$-$\ms$ relationship for BCGs found by \cite{McConnell}, to arrive at a density of $6\times 10^5 \msun/{\rm Mpc}^3$.
Due to the observed constancy of the $\mbh$-$\ms$ relationship for BCGs, and the small fractional change in density from accretion for massive
black holes \citep{Yu}, we are justified in treating $\langle\rbh \rangle$ as constant.
 
Finally, we solve for the redshift dependence of $\al$ that satisfies mass conservation.
We expand the resulting $\al$ to $\mathcal{O}(1/(1+z))$, and find that
\beq
\al \approx -2.0+\frac{0.52}{1+z}\,,
\eeq
is accurate to within a percent for $0\leq z\leq 1$.
We note that the redshift dependence of $\al$ agrees reasonably well with the literature (see \eg \cite{MOIRCS}); however,
a constant $\nc$ is generally inconsistent with observational data in the literature, 
which typically measures $\nc$ to be a decreasing function of redshift.
We suspect sampling effects may account for this to some degree, with galaxies 
evolving from blue to red, and effectively ``appearing'' in a survey at lower redshifts,
thereby increasing the apparent number density of massive red galaxies.  In contrast, our approach conserves, by construction,
the mass density of old stars.

The principle theoretical reason for choosing a constant $\nc$ is that mergers involving massive galaxies are unlikely to be 
equal mass mergers, given the rarity of black holes with $M_{\bullet}\sim\mbhc$.
To intuit why this implies a constant $\nc$, and what behaviors we might expect for $\mc$ and $\al$, we can use a very simple model of galaxy evolution
solely for the purpose of gaining intuition about the evolution of the Schechter parameters.
Instead of using linear-in-mass bins for the mass function, 
we can imagine that mass-doubling occurs only through equal mass mergers,
whereas mass doubling of high-mass bins occurs through the depletion of the lowest-mass bins in the
minor-merger scenario.  For the major-merger scenario, after doubling, every log-mass bin will simply shift
toward larger mass by
one log-mass unit and will halve their number density, 
doubling $\mc$, halving $\nc$, and leaving $\al$ constant.  However, for the minor-merger scenario,
the number of very massive galaxies remains fixed while they accrete smaller satellites, so that after mass doubling, 
massive number-density bins will simply shift by one log-mass unit toward higher mass at fixed number density,
whereas low-mass bins will shift toward higher mass with their number density being depleted by mergers with more massive galaxies, and replenished by mergers between lower mass members.
If we require that the distribution maintains a single power-law mass distribution for less massive galaxies, then in the minor-merger scenario 
the low-mass bins must have their number density depleted 
relative to the high-mass bins, in which case the slope of the mass function at low masses
would evolve toward shallower, less negative $\al$ values as we evolve toward lower redshift.

\subsection{Evolution of the Mass Function}
\label{sec:ev}
\begin{figure}
\includegraphics[trim = 0mm 0mm 0mm 0mm, width=0.45\textwidth]{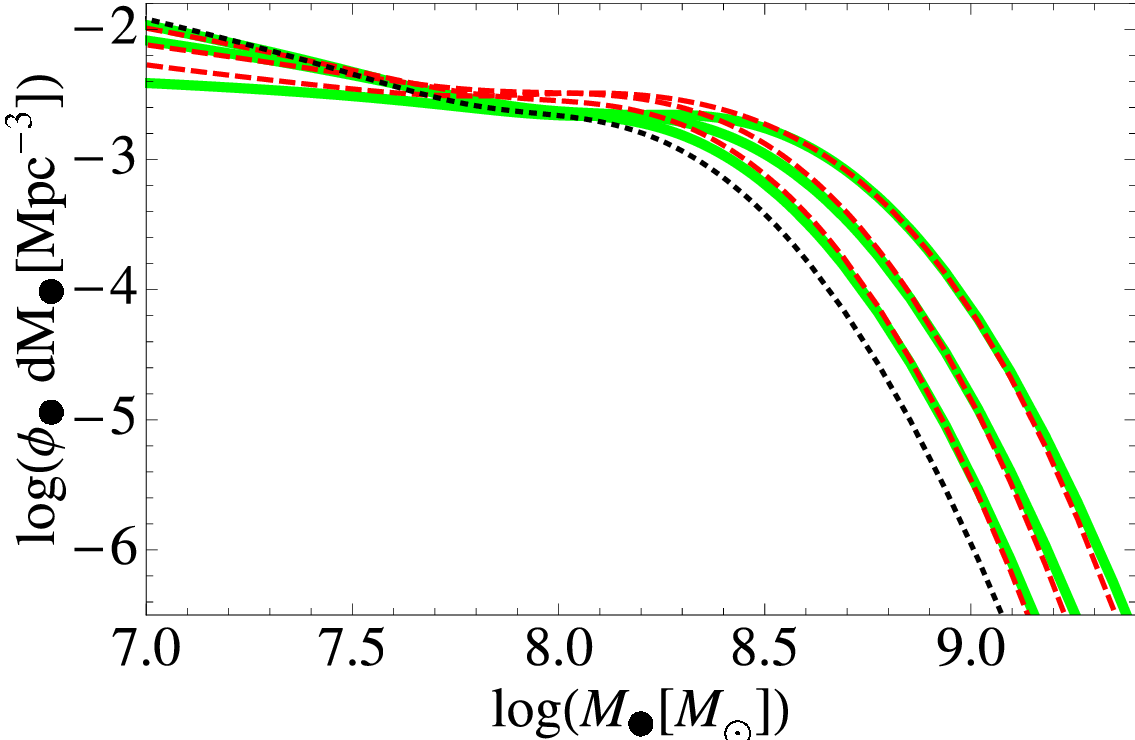}
\hfill
\caption{Evolution of the black-hole mass function (equivalently, the stellar mass function using $\ms=625\,\mbh$) from $z=1$ to $z=0$.
The dotted (black) curve shows the initial $z=1$ mass function, and the solid (green) and dashed (red) curves show the analytic (described
in Sec.~\ref{sec:sch})
and evolved (described in Sec.~\ref{sec:ev}) mass functions, respectively, at $z=\{2/3,\,1/3,\,0\}$.
\label{fig:nvM}
}
\end{figure}

In this section, we investigate how closely the observed evolution with redshift of the mass function
can be reproduced with a simple model that assumes merger-driven evolution over the redshift interval $z\leq 1$.  The evolution
of the mass function generally results from some combination of mergers, star formation, and mass loss (\eg through winds):
\beq
\frac{\partial}{\partial z}\left(\phi\, dM\right) dz = \left[\frac{\partial}{\partial z}\left(\phi\, dM\right) dz\right]_{\rm m} - \frac{\partial}{\partial M}\left(\phi\, \bigg\langle \frac{\partial M}{\partial z}\bigg\rangle \right) dM dz\,,
\eeq
where $\langle \frac{\partial M}{\partial z}\rangle \approx \frac{\partial \mc}{\partial z}$ is the mean rate of change of mass at
fixed number densities.
To find the contribution from mergers, $\left[\frac{\partial}{\partial z}\left(\phi\, dM\right) dz\right]_{\rm m}$, we need to know the number of mergers between a black hole with $\mc$-normalized mass between $M'$ and $M'+dM'$ and another
with normalized mass between $M''$ and $M''+dM''$ at a redshift between $z$ and $z+dz$.  This requires some care, since BCGs will not
merge with each other, so they must be handled separately.  Since we are assuming merger-dominated evolution, the mass dependence of
the merger probability distribution will be proportional to the number densities of the two merging black holes.  The redshift dependence 
$P(z)$ can be chosen to maximize the agreement between between our model and the observed evolution of the mass function.  We therefore
express the number density of mergers between black holes of $\mc$-normalized mass $M'\equiv M_{\bullet}'/\mc_{\rm r}$ and 
$M''\equiv M_{\bullet}'/\mc_{\rm r}$ at redshift $z$ (where $\mc_{\rm r} \equiv (1+z)\mc$ is the redshifted characteristic black hole mass) as 
\begin{widetext}
\beq
\frac{\partial^3\phi_{\{\rm low,\,BCG\}}}{\partial M' \partial M'' \partial z} dM' dM'' dz = P(z) dz\, \phi_{\{\rm tot,\,BCG\}}(M')\, dM' \,\phi_{\{\rm tot,\,low\}}(M'') \,dM''\,,
\label{eq:bivar}
\eeq
\end{widetext}
where in the notation $\phi_{\{\rm a,\,b\}}$, the first arguments all form an equation for $\phi_{\rm a}$, and the second arguments
form an equation for $\phi_{\rm b}$.  In addition, $P(z)$ can be separated into a purely cosmological component
and a comoving merger probability component as $P(z) \equiv \frac{dp}{dV_c}\frac{dV_c/dz}{n(z)}$, 
$\frac{dp}{dV_c}$ is the merger probability in a fixed
comoving volume, $n(z)$ 
is the total number density at a given redshift, and $dV_c/dz$ accounts for cosmological expansion and is given by
\beq
\frac{dV_c}{dz} = \frac{4\,\pi \,c\,D_L^2}{H_o(1+z)^2\sqrt{\Omega_M(1+z)^3+\Omega_{\Lambda}}}\,.
\eeq
We assume a flat cosmology with WMAP seven-year values for the cosmological parameters \citep{WMAP7}; therefore, the luminosity distance is given by
\beq
D_L(z)=\frac{c(1+z)}{H_o} \dint_{0}^{z} \frac{dz'}{\sqrt{\Omega_M(1+z')^3+\Omega_{\Lambda}}}
\eeq
(see \cite{Hogg} and references therein).
We can then find the number of black holes that leave (the sink term ``si'') and enter (the source term ``so'')
a mass bin at $\mbh$ during the interval $dz$,
\begin{widetext}
\bea
\left[\frac{\partial}{\partial z}\phi_{\{\rm low,\,BCG\}}\,(\hat{M})\right]_{\rm si} &=& P(z)\dint_{\mlow /\mc_{\rm r}}^{\mhi /\mc_{\rm r}} \phi_{\{\rm tot,\,BCG\}}\,(\hat{M}) \phi_{\{\rm tot,\,low\}}(M')\, dM' \nonumber \\
\left[\frac{\partial}{\partial z}\phi_{\{\rm low,\,BCG\}}\,(\hat{M})\right]_{\rm so} &=& P(z)\dint_{\mlow /\mc_{\rm r}}^{(\hat{M}-\mlow)/\mc_{\rm r}} \dint_{\mlow /\mc_{\rm r}}^{(\hat{M}-\mlow )/\mc_{\rm r}} \delta(\hat{M}-M'-M'') \phi_{\{\rm low,\,BCG\}}(M') \,dM' \phi_{\{\rm low,\,low\}}(M'') \,dM'' \nonumber \\
&=& P(z)\dint_{\mlow /\mc_{\rm r}}^{(\hat{M}-\mlow) /\mc_{\rm r}} \phi_{\{\rm low,\,BCG\}}(M')\, \phi_{\{\rm low,\,low\}}(\hat{M}-M') \,dM'\,.
\label{eq:siso}
\eea
\end{widetext}
We will consider the mass range $\mlow \leq M_{\bullet} \leq \mhi$ throughout this work, although the precise choices do not 
significantly affect any of the results that we present.
All that remains is to 
determine whether $\frac{dp}{dV_c}$ can be well-approximated (and
replaced in our evolution equation) by a constant value.
We find that fixing $\frac{dp}{dV_c}=260\,{\rm Mpc}^3$ in our evolution equation reproduces the redshift-dependence of
the mass function using our redshift-dependent Schechter parameters remarkably well (see Fig.~\ref{fig:nvM}).

It is noteworthy that, despite the good agreement between our model based on galaxy evolution through mergers and our adopted Schechter-based mass function, the evolved solutions
can only approximate Eq.~\ref{eq:mf}.
The analytic form of the source term involves generalized hypergeometric functions among others, so that the evolved 
mass function does not simplify to the functional form of Eq.~\ref{eq:mf}.
The shape of the transition from less massive black holes to the BCG component
as well as the exponential decay for very large masses in the evolved model 
depend sensitively on the location of the break mass $\mathcal{M}$, which sets 
both the ``knee'' location and the e-folding, so a sufficiently fine redshift sampling (or continuous integration, as we have done)
is necessary to accurately approximate the target mass function at each redshift.

\section{Satellite black hole dynamics within the host galaxy}
\label{sec:dyn}
Having established a method for calculating the distribution of galaxy mergers, we must now determine whether those mergers lead to
the merger of the concomitant SMBHs and thereby contribute to the gravitational wave background.
In the case of major mergers, the stellar bulges and their central black holes will usually merge in less than a Hubble time.  We loosely define 
major mergers as having mass ratios $q\equiv M^s/M^h\geq 1/4$, where $M^h$ is the larger ``host'' mass, and $M_s$ is the ``satellite'' mass (either
black hole or bulge).
For minor mergers, the satellite black hole will likely be 
stripped of its stellar bulge, which is deposited outside the host bulge and explains the exceedingly large increase
in the apparent size of galaxies since $z=1$ \citep{vanDokkum} in this model.  
However, the satellite black hole must be able to migrate to the central core of the host under dynamical friction
by the present day to contribute to the gravitational wave signal.  To determine whether this occurs, 
we use the well-known expression for the dynamical
friction timescale \citep{Chandrasekhar:1943ys,TOS} in a convenient form (Eq. 8.12 in \cite{Binney}),
\bea
t_{\rm DF} &=& \frac{19 {\rm Gyr}}{\ln (1+M^h_{\star}/M^s_{\bullet})} \left(\frac{R_e}{5 {\rm kpc}}\right)^2 \frac{\sigma (R_e)}{200 {\rm km/s}} \frac{10^8 \msun}{M_s} \nonumber \\
&\approx& \frac{4.5 {\rm Gyr}}{q(6.9-\ln q)} \left(\frac{M^h_{\bullet}}{10^8\msun}\right)^{2/3} (1+z)^{-3/2}\,,
\label{eq:tdf}
\eea
where $R_e$ is the half-light radius of the host galaxy and $\sigma$ is the local velocity dispersion, for which we use
\bea
R_e &=& 2.5\,{\rm kpc} \left(\frac{\mbh}{10^8\,\msun}\right)^{0.73}(1+z)^{-1.44}\,{\rm and} \nonumber \\
\sigma (R_e) &=& 190\,{\rm km/s} \left(\frac{\mbh}{10^8\,\msun}\right)^{0.2}(1+z)^{0.44}\,,
\label{eq:re}
\eea
where the mass-dependence comes from fits to Sloan Digital Sky Survey (SDSS) data \citep{Nipoti}, and the redshift dependence comes from fits
to simulation results which were shown to be consistent with various surveys in \cite{Oser} within the redshift range we consider.  

We note that Eq.~\eqref{eq:tdf} differs in multiple ways from what is often used in merger-tree models (see Eq.~9 in \cite{Volonteri}).
Specifically, it is often assumed that $t_{\rm DF}$ depends only on the mass ratio of the merging pair, \ie dynamical friction
does not depend on the mass scale.  This would be a very surprising result for multiple reasons.  The appropriate $t_{\rm DF}$ for merger-trees
should incorporate both the merging dark matter halos and, subsequently, the stellar bulges.  The mass
of the host and the inclusion or exclusion of its stellar bulge will affect its equilibrium density distribution,
and will therefore affect the tidal stripping of the satellite.  Even if the inclusion of the bulge does not affect
$t_{\rm DF}$ significantly as argued in \cite{Boylan}, the radius of injection for the satellite and the
orbital kinematics of the satellite should depend directly on the host mass.
Since we need an expression for $t_{\rm DF}$ that applies for
$t_{\rm DF}\approx t_{\rm H}$ in order to determine whether the host and satellite will merge, 
the scenario of a ``naked'' satellite black hole moving through the host bulge by dynamical
fraction is precisely the case of interest.  Therefore, in Eq.~\eqref{eq:tdf}, the relevant mass ratio is $M^h_{\star}/M^s_{\bullet}$,
rather than $M^h_{\star}/M^s_{\star}$ as is often used in dark matter merger trees, such
as the model of \cite{Volonteri} which has been employed frequently for gravitational wave 
calculations.  
Finally, in the scenario of a stripped satellite, the satellite black hole is effectively deposited at $R_e$ within the
host galaxy, and spirals toward
the center through dynamical friction.  $R_e$ is
far smaller than the virial radius, which is the fiducial distance scale used in dark-matter merger-tree simulations \citep{Volonteri}, 
where the merger of two halos is considered,
and the bulges are ignored when calculating $t_{\rm DF}$.
 
At each redshift, we use Eq.~\eqref{eq:tdf} to determine which black holes will merge.
In order to calculate the total gravitational wave signal from the merging population using the Monte Carlo method, we need to sample from the probability distribution $\frac{d^4p}{dM_1dM_2dz\,df}$.
Eq.~\ref{eq:bivar} provides most of what we need, since $\frac{d^3p}{dM_1dM_2dz}=\frac{1}{n(z)}\frac{d^3n}{dM_1dM_2dz}$.  The probability of finding a binary in a given frequency
interval $df$ is found by observing that the quadrupole frequency rate $\dot{f}\propto f^{11/3}$, so that $\frac{dp}{df}\propto f^{-11/3}$, with the normalization being determined
by the range of possible frequencies.  

It is known that dynamical friction becomes ineffective once the binary reaches a separation of $\mathcal{O}$(1 pc),
so we must assume a mechanism for solving the ``last parsec problem'' \citep{Merritt:2005LR} in order to determine the minimum
frequency where gravitational radiation will drive the evolution of the binary.  Following \cite{Quinlan96},
we assume the binary hardens through the repeated scattering of stars in the core of the host, until gravitational radiation
becomes the dominant process at
\beq
f_{\rm min}=2.7 \times 10^{-9}\,{\rm Hz} \left(\frac{M^h_{\bullet} M^s_{\bullet}}{(10^8\,\msun)^2}\right)^{-0.3}\left(\frac{M^h_{\bullet}+M^s_{\bullet}}{2\times 10^8\,\msun}\right)^{0.2}.
\label{eq:fmin}
\eeq
The maximum frequency is set to the innermost stable circular orbit (ISCO) for a test particle orbiting a Schwarzschild black hole
with the same total mass,
\beq
f_{\rm max}=f_{\rm ISCO}=1.1\times 10^{-4}\,{\rm Hz} \left(\frac{M^h_{\bullet}+M^s_{\bullet}}{2\times 10^8\,\msun}\right)^{-1}\,,
\eeq
though we emphasize that the results are insensitive to the precise choice of $f_{\rm max}$, since we perform a random Monte Carlo
draw from an $f^{-11/3}$ distribution, so only $f_{\rm min}$ plays a critical role.
Using a Monte Carlo sample to set the starting frequency, we invert
\beq
f(t)=1.4\times 10^{-6}\,{\rm Hz} \left(\frac{(10^8\,\msun)^2}{M^h_{\bullet} M^s_{\bullet}}\frac{10\,{\rm yr}}{t_m-t}\right)^{3/8}\left(\frac{M^h_{\bullet}+M^s_{\bullet}}{2\times 10^8\,\msun}\right)^{1/8}\,,
\label{eq:f}
\eeq
to find the corresponding time before merger, then reapply Eq.~\eqref{eq:f} to find the frequency after a $T=5$ year observation, where $t_m$ approximates the time of the final merger.
We note that all of the frequency expressions given above are appropriate for the rest frame of the source, $f_r$, and must be redshifted 
to give the measured gravitational wave frequency, $f_m$, using $f_m=f_r/(1+z)$.  With the range of permitted frequencies now determined by the dynamics of each binary,
we can generate gravitational wave signals from each, and combine them to form our estimate of the stochastic gravitational wave background. 

\section{Gravitational waves from SMBH mergers since $z=1$}
\label{sec:gw}
In order to calculate the total gravitational wave signal from all SMBHBs in the PTA band, we must combine our knowledge of the merger probability
distribution $\frac{d^3p}{dM_1dM_2dz}$ and the total number of mergers $N$ with the signal strength of each source, given in units of dimensionless strain $h$.
If the resulting span of frequencies $\Delta f \equiv f(t)-f(t-T) < T^{-1}$ for a given source, then we assign a gravitational wave strain amplitude appropriate
for a monochromatic source, which is given within the quadrupole approximation by Eq.~58 of \cite{Thorne300},
\bea
h &=&3.2\times 10^{-17}\, \left(\frac{M^h_{\bullet} M^s_{\bullet}}{(10^8\,\msun)^2}\right) \nonumber \\
&\times& \left(\frac{2\times 10^8\,\msun}{M^h_{\bullet}+M^s_{\bullet}}\right)^{1/3}\left(\frac{1\,{\rm Gpc}}{D_L}\right)\left(\frac{f}{10^{-7}\,{\rm Hz}}\right)^{2/3}\,,
\label{eq:hm}
\eea
where this amplitude is already averaged over source polarization.  Sources with $\Delta f > T^{-1}$ are potentially resolvable, and are not part of the stochastic background,
since they introduce a correlation in the signal between adjacent frequency bins.  In addition, sufficiently nearby sources may be sufficiently bright to be resolved even if they are
monochromatic, but these sources would also need to be removed before the data could be considered stochastic.  Therefore, we exclude both of these types of 
sources from the current study.  We note that these resolvable sources, although much rarer than sources contributing to the stochastic signal, will nonetheless dominate
the total energy density in gravitational waves, and are the only means by which we might gain knowledge about individual binary systems.  Resolvable sources are therefore of great
interest as well; however, since they require a much larger and more computationally expensive Monte Carlo sampling due to their rarity, we leave them for future work.

Finally, we calculate the full stochastic gravitational wave signal by summing the power of each source, thereby approximating
\beq
h_{\rm c}^2(f)=\dint_{0}^{1}dz \dint_{\mlow}^{\mhi}dM_2 \dint_{\mlow}^{M_2} dM_1 Nh^2 \frac{d^4p}{dM_1dM_2dz\,df}\,,
\eeq
where 
\beq
N\equiv \dint_{0}^{1}dz \dint_{\mlow}^{\mhi}dM \left[\frac{\partial}{\partial z}\phi \right]_{\rm so}
\label{eq:num}
\eeq
is the total number of mergers during the full observation, and the frequency minimum and resolution are set assuming a 5 year observation.  
Since we have shown that our evolution duplicates the observed mass function over the mass-range of interest, we 
draw merging pairs from the probability distribution formed using the analytical fit to the observed mass function at each redshift, 
rather than from the numerical mass function generated through our evolution.
Due to the extremely large number, $\mathcal{O}(10^{11})$, of mergers required to duplicate the observed evolution of the mass function,
and the fact that low mass black holes cannot be excluded \emph{a priori} since they dominate the overall number density,
we have performed a Monte Carlo for $N_{\rm MC}=5\times 10^6$ binaries, and scaled the result appropriately.  To scale, we first fit the Monte Carlo result
to the form 
\beq
h_{\rm fit}\equiv h_{\rm o} \exp\left[\left(\frac{f}{f_{\rm DF}}\right)^{\beta}\right] \left(\frac{f}{f_{\rm o}}\right)^{-2/3}\left(1+\frac{f}{f_{\rm o}}\right)^{\gamma}\,,
\label{eq:fit}
\eeq
where $h_{\rm o}$, $f_{\rm DF}$, $f_{\rm o}$, $\beta$, and $\gamma$ are tunable parameters.
 The $f^{-2/3}$ term is the usual spectrum for stochastic binaries \citep{Phinney,Jaffe}. The final term in Eq.~\eqref{eq:fit} is the same one used
in \cite{Sesana:2008mz} and accounts for the discrete nature of the sources and its effect at high frequencies where the number of sources diminishes.
The exponential term is introduced here for the first time, and is primarily motivated by the suppression of high mass sources due to dynamical
friction (i.~e.~the $M^h_{\bullet}$-dependence of $t_{\rm DF}$ in Eq.~\eqref{eq:tdf}), which was also observed in \cite{Kocsis}.  We note
that the mass dependence of $f_{\rm min}$ in Eq.~\eqref{eq:fmin} can also lead to biasing against massive merging systems, so that the
mechanism for solving the last parsec problem can also influence this region of the spectrum.
We will use the parameterization of Eq.~\ref{eq:fit} to first fit, then scale the results from our Monte Carlo simulation.

After calculating the best-fit parameters from Eq.~\eqref{eq:fit}, we multiply the resulting $f_{\rm o}$ by
$\left(N/N_{\rm MC}\right)^{3/11}$ (since $N\propto dp/df \propto f^{-11/3}$ as previously mentioned), and we multiply
the overall amplitude of $h_{\rm fit}$ by $\sqrt{N/N_{\rm MC}}$ (equivalently, we multiply $h_{\rm o}$ by $\left(N/N_{\rm MC}\right)^{7/22}$,
since $h_{\rm fit}\propto h_{\rm o}\,f_{\rm o}^{2/3}$ for $f \ll f_{\rm o}$).
We find the lowest frequency bin at which the Monte Carlo
simulation only predicts a single source, we again multiply by $\left(N/N_{\rm MC}\right)^{3/11}$, and we terminate the scaled fit at that frequency,
since the signal is clearly no longer stochastic.  The parameters $f_{\rm DF}$, $\beta$, and $\gamma$ remain fixed in our rescaling, since
there is no apparent physical grounds for these parameters to depend on $N$, and empirically we find that fixing these values from one
Monte Carlo provides a good fit for other simulations with different values of $N$.

\begin{figure}
\includegraphics[trim = 0mm 0mm 0mm 0mm, width=0.48\textwidth]{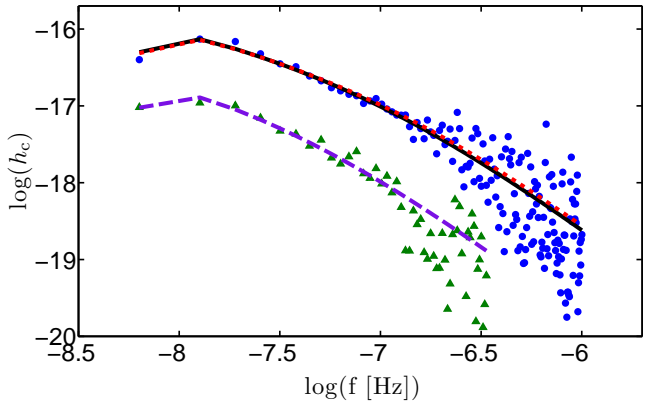}
\hfill
\caption{Gravitational wave strain demonstrating the effectiveness of scaling our Monte Carlo results to cases with a larger number of sources.
The (green) triangular markers show the result of a Monte Carlo where $N_{\rm MC}=10^5$, and the (blue) circular markers
show another Monte Carlo with $N_{\rm MC}=5\times 10^6$.  We fit the signal from the smaller sample
to Eq.~\ref{eq:fit}, resulting in the (purple) dashed line, then scale the resulting amplitude, reference frequency, and final frequency to $N=5\times 10^6$ as discussed in the text.  The result is the
(red) dotted curve, which provides a very accurate representation of the Monte Carlo result from the larger sample.  To demonstrate this, we also fit the signal from the larger sample using
Eq.~\ref{eq:fit}, with the (black) solid line being the result, which agrees well with the scaled fit to the signal from the smaller Monte Carlo sample.
\label{fig:scale}
}
\end{figure}

Fig.~\ref{fig:scale} demonstrates the effectiveness of the overall scaling procedure by comparing two different Monte Carlo simulations.
One simulation has $N_{\rm MC} = 5\times 10^6$, and the other has $N_{\rm MC} = 10^5$.  We show the raw strain data from both simulations,
along with the best fits of each simulation using Eq.~\eqref{eq:fit}.  We then scale the simulation with $N_{\rm MC} = 10^5$ according
to the scaling rules we have described, in order to compare with the actual fit of the $N_{\rm MC} = 5\times 10^6$ simulation.
Despite the scatter in the data, particularly at large frequencies, the scaled fit agrees extremely well with the larger Monte Carlo fit,
indicating that the scaling of the mean gravitational-wave strain is very robust.

\begin{figure*}
\begin{center}
\includegraphics*[trim = 0mm 0mm 0mm 0mm, width=0.98\textwidth]{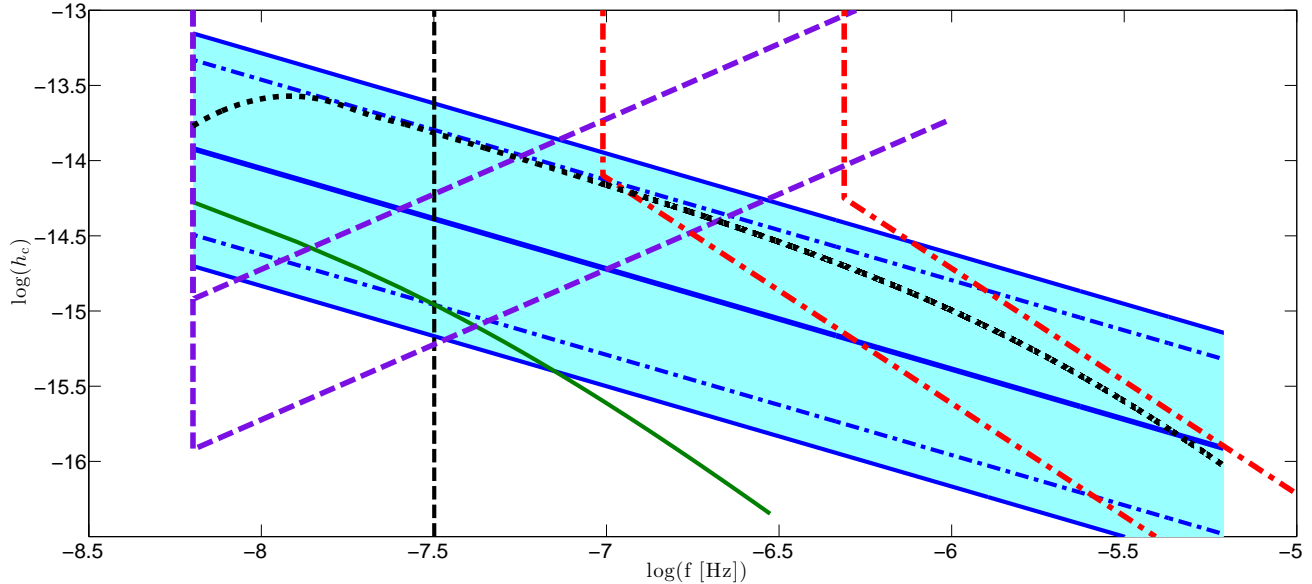}
\hfill
\caption{Gravitational wave strain and strain sensitivities for a 5 year integration.  The strain signals are orientation-averaged, and the sensitivities are sky-averaged.
The dotted (black) line shows the scaled Monte Carlo results for the stochastic signal from merging massive black holes at $z\leq 1$ using our fiducial mass function parameters,
and the (cyan) shaded region shows our estimate of the possible range of stochastic signals based on uncertainties in our model and assuming
a conventional $f^{-2/3}$ spectrum, with the median signal shown as a thick solid (blue) line, and the optimistic and conservative \{2$\sigma$, 3$\sigma$\} limits 
shown with thin \{dash-dotted, solid\} (blue) lines (see the text for details on our approximation of the uncertainty intervals).
The thin solid (green) curved spectral line shows the signal predicted in \cite{Sesana:2008mz}, which is $\sim 4\times$ weaker than our median estimate at $f=1$ yr$^{-1}$.  The PTA sensitivities are
scaled to match the 95\% confidence limit of current PTAs and a future instrument like the SKA with $10\times$ better timing accuracy, where the current and future sensitivities
are shown as dashed (purple) lines with lesser and greater sensitivity, respectively, and the typical PTA reference frequency
of $f=1$ yr$^{-1}$ is shown as a vertical thin dashed (black) line.  We note that a detectable signal (at 95\% confidence) will intersect the PTA limit at $f=1$ yr$^{-1}$, indicating
that much of the parameter space for the mass function has already been excluded in our model, and that a likely detection in the near future is an overall prediction of our model.
Finally, the dash-dotted (red) lines show an optimistic
estimate of LISA's and SGO's low frequency sensitivity (less and more sensitive, respectively), which indicates that SMBHB mergers at $z\leq 1$ may be
a potential source of detectable stochastic gravitational waves below $f\sim 10^{-5}$ Hz for space-based gravitational wave detectors.
\vspace{0.15in}
\label{fig:h}
}
\end{center}
\end{figure*}

In Fig.~\ref{fig:h}, a Monte Carlo with $N_{\rm MC}=5\times 10^6$ scaled
to the actual number of expected binary sources ($N=4.75\times 10^{11}$ using Eq.~\ref{eq:num}) is shown,
where the scaled-fit values (\ie our estimate for $h_{\rm c}$ that we would expect to observe from massive merging binaries at $z\leq 1$) are
$h_{\rm o}=8.4\times 10^{-16}$, $f_{\rm o}=2.5\times 10^{-6}$ Hz, $f_{\rm DF}=6.3\times 10^{-9}$ Hz, $\gamma=-1.3$, and $\beta =-4$,
with the final frequency cutoff occurring at $6.2\times 10^{-6}$ Hz.
We also estimated our uncertainty in $h_{\rm c}$ as follows.
We first assumed a $\pm 10$ \% standard deviation in $\log (\mc )$, $\nc$, $\log (\rbh )$, and $\sigma$, consistent with
experimental ranges \citep{Drory,MOIRCS}. We then ran Monte Carlo simulations for all 8 variations, and calculated
best-fit parameters for $h_{\rm c}$ using Eq.~\eqref{eq:fit}.  However, since all analyses of PTA data to-date have
assumed a simplified form for the spectrum of the stochastic background, which we label $h_{\rm s}$ and which is given by \citep{Jaffe}
\beq
h_{\rm s} \equiv h_{\rm o} \left(\frac{f}{{\rm yr}^{-1}}\right)^{-2/3}\,,
\label{eq:hs}
\eeq
we must convert our range of $h_{\rm fit}$ functions to a range of functions of the form given by
Eq.~\eqref{eq:hs}.  We therefore approximate the expected and lower-limit strain spectra by requiring that $h_{\rm s} \leq \langle h_{\rm fit} \rangle$
and $h_{\rm s} \leq {\rm min}(h_{\rm fit})$, respectively,
over the full band of the stochastic signals, using all 17 Monte Carlo realizations (one central parameter set, and 16 more from perturbing all four Schechter parameters
up and down at both the 2$\sigma$ and 3$\sigma$ levels).  We likewise require that $h_{\rm s} \geq {\rm max}(h_{\rm fit})$ for the upper-limit strain spectrum.  In this way,
the uncertainty region shown in Fig.~\ref{fig:h} encompasses the full range of spectra resulting from the 2$\sigma$ and 3$\sigma$ confidence
interval of best-fit Schechter functions to observational data.  Because $h_{\rm fit}$ is the trend of the Monte Carlo results by construction,
the fluctuations around the trend evident in Fig.~\ref{fig:scale} are not included in the uncertainty interval estimate shown in Fig.~\ref{fig:h}.
Given the large number of total sources, and their concentration at $f \leq {\rm yr}^{-1}$, we do not expect the variance $\sigma^2$ of the spectra to
significantly affect the constraint for the particular Monte Carlo realization produced by nature, since $\sigma^2 (f) \approx h_{\rm c} (f)/N(f)$.

We must emphasize that the procedure we have just described is only a rough estimate of how our model should translate to confidence
intervals derived assuming a simple $f^{2/3}$ spectrum.  Firstly, the observational limits can only be used to imply detectability thresholds, as we have done,
if the limits are derived from the same statistic (namely, the cross-correlation of signals from different pulsars) that must be used for detection.  If, for instance,
a tighter constraint is found through the use of autocorrelation of the signals from individual pulsars, we cannot assume that this constraint implies detectability
of a signal at that level.  This is not an issue for the constraints we quote, but may become an issue with future constraints.  Secondly, it is important to bear in mind that
any constraint assuming a simple $f^{2/3}$ spectrum will be dominated by the signal content at $f\approx T_{\rm obs}^{-1}$, the inverse observation time.
However, as is clear from the form of $h_{\rm fit}$ and from Fig.~\ref{fig:h} that, at least for some choices of Schechter parameters, the signal strength becomes diminished at
the lowest frequencies, due to the failure of sufficiently massive hosts to consume any satellites through dynamical friction quickly enough.  Therefore, the detectability
of the signal will depend very sensitively on the behavior of the actual spectral shape of the signal at $f\sim T_{\rm obs}^{-1}$ that is realized in nature.  For cases
where the signal does dip, the sensitivity will no longer accumulate as $T_{\rm obs}^{13/6}$ \cite{Jenet2005}, but instead will accumulate like the SNR of the most sensitive frequency,
which will be fixed and independent of $T_{\rm obs}$.

In addition to our estimate for $h_{\rm c}$, we show the results of \cite{Sesana:2008mz}, which agrees with \cite{Sesana2010}.
The large discrepancy between these results and our expected strain spectra (our mean estimate is $3.3\times$ larger at $f= 0.2$ yr$^{-1}$ and
$5.1\times$ larger at yr$^{-1}$) requires an explanation, so we will address this disagreement before we discuss the remainder of Fig.~\ref{fig:h}.\footnote{We note that
the recent results in \cite{Sesananote} are somewhat less in conflict with our constraints, differing only by a factor $\sim 2.5$.  In this case,
a lower implied merger rate may explain a larger fraction of the overall difference.}
\cite{Sesana:2008mz} construct $\frac{d^3n}{dM_1dM_2dz}$ using the methods described in \cite{Volonteri}; specifically, the extended
Press-Schechter formalism \cite{Bond} is used to construct a dark-matter-halo merger tree, which is populated with black holes using the $\mbh-\sigma_{\star}$
or $\mbh-\ms$ relationships, with the baryon dynamics treated as we described in our discussion of dynamical friction and the last parsec problem.
\cite{Sesana:2008mz} also employ other models than that in \cite{Volonteri}, but these models differ in ways that are not relevant to our
discussion (such as details in the way that feedback mechanisms are included), and the results from all of the models considered in \cite{Sesana:2008mz}
agree with each other far more than they agree with our results.
\cite{Sesana2010} use a different method altogether, yet arrive at a nearly identical result.  They use $\frac{d^3n}{dM_1dM_2dz}$ by populating
the catalog of merged galaxies constructed in \cite{Bertone} with black holes based again on the $\mbh-\sigma_{\star}$ or $\mbh-\ms$ relationships.
\cite{Bertone} used the Millennium simulation \citep{Millenium} to determine the merger rates of dark matter halos, and used semi-analytic
techniques to approximate detailed baryonic physics such as star formation, accretion, stellar and supernova wind feedback, and the effects of metallicity.
However, while \cite{Bertone} do not explain in detail their treatment of baryon interactions during galaxy mergers, they do state that the
``outcome'' of a galaxy merger depends only on the mass ratio in their model, and no mention is ever made of black holes stripped of their stellar bulge.  Therefore,
we conclude that the treatment of dynamical friction and the last parsec problem is qualitatively no different from that used in \cite{Volonteri},
and therefore suffers from the same potential shortcomings for cases where the satellite black hole is stripped of its bulge.  Since this is seen to occur in galaxy merger
simulations of massive galaxies \cite{Bellovary}, we feel this is a likely source of discrepancy between our result and past estimates.

In addition to issues with the treatment of satellite dynamics within the host galaxy, two other factors that are
shared by the treatments in \cite{Sesana:2008mz} and \cite{Sesana2010} contribute to their disagreement with our result; the first
issue relates to the initial conditions of the dark matter simulations that are used, and the second issue arises from the nontrivial relationship between halo merger rates
and galaxy merger rates.  
The Millennium simulation used by \cite{Sesana2010} took the observed fluctuations in the power spectrum from the first year of WMAP data \cite{WMAP1}
and used the data to determine the initial distribution of dark matter particles at $z=127$.  The model presented in \cite{Volonteri} and used in
\cite{Sesana:2008mz} predates WMAP, so that the abundance of X-ray emitting
clusters in the local universe as measured in \cite{Eke} was used to normalize a theoretical fit to the CMB power spectrum \citep{Bardeen86,Sugiyama}.
In both cases, the amount of power contained in high-$\ell$ multipole moments/small angular scales was underestimated in comparison to results
from the completed seven year WMAP survey \citep{WMAP7}.  Since the black holes that we are interested in observing affect the power spectrum
on small angular scales, this could have a significant impact on the total number of black holes, and therefore the number of mergers, that
these models would predict.  

Finally, we consider the relative merger rates of dark matter halos with and without cores of baryonic matter (i.~e.~the stellar component of the galaxy).
From Eq.~\eqref{eq:tdf}, we see that the dynamical friction timescale $t_{\rm df} \propto M^h/M^s$.  If a satellite halo is spiraling through a large host halo,
as would be the case in a galaxy cluster, then the time it will take for the satellite to spiral to the center of the cluster scales inversely with $M^s$.  The size and mass
of the satellite halo will be limited by tidal stripping, which will be more severe in the absence of a stellar core.  The remaining halo,
though still larger than the stellar core, is far less dense, so the core may dominate the total satellite halo mass.  Therefore, the inclusion of the stellar core could
greatly decrease $t_{\rm df}$ and thereby increase the merger rate.  By using the observed evolution of galaxies to calibrate $\frac{d^3n}{dM_1dM_2dz}$,
our model automatically includes the correct baryonic physics, whereas including the correct physics semi-analytically in dark matter simulations is far from trivial.
Therefore, the primary cause for the difference between our merger rate and the rate found in prior publications is that we estimate the merger rate using the observed
density of massive galaxies, whereas most prior estimates have been based on dark-matter merger trees.
 
Returning to our description of Fig.~\ref{fig:h}, we also show the rms strain sensitivity $h_{\rm rms}$, averaged over sky location, 
for the current European/NANOGrav/Parkes PTA (using the European constraint and assuming 100 ns timing accuracy for all pulsars), 
and the Square Kilometer Array (10 ns accuracy), assuming all arrays
observe 20 pulsars for 5 years (see \cite{Sesana:2008mz} for details on calculating PTA sensitivities).  
Since a significant fraction of the overall range of our signal estimates is above the current PTA sensitivity
at $f=$yr$^{-1}$, it is likely that a rigorous
analysis of actual PTA data using a model that better represents the behavior of the Monte Carlo results will already exclude large regions of the
parameter space of Schechter parameters, assuming the overall validity of our merger-dominated assumption.  Furthermore, our most
conservative estimated level for the stochastic background would provide a detection with the SKA monitoring 20 pulsars for 5 years
at a far greater statistical significance than the expected strain level from \cite{Sesana:2008mz,Sesana2010}.  If our model is valid, the SKA
should not be necessary for a detection, as continued observation of the current set of pulsars with the current level of timing accuracy
has a high likelihood of yielding a detection long before the SKA begins collecting data.
 
We also show the rms sensitivity for both the original Laser Interferometer Space Antenna (LISA) design, and a revised lower cost 
design (SGO Mid) currently under consideration by NASA, both calculated
using \cite{LISASenGen} with the specifications for each design found in \cite{RFI}
(a design with near-identical sensitivity called NGO is also being considered by the European Space Agency).  
We show sensitivity estimates calculated
directly from \cite{LISASenGen}, although we note that the sensitivity of these designs below $\sim 3\times 10^{-5}$ Hz is highly uncertain,
and the results from \cite{LISASenGen} are likely a best-case scenario.
In addition to the stronger predicted signal within the PTA band, we predict the potential for significant signal within the LISA/SGO band as well,
if the lowest frequency sensitivities from \cite{LISASenGen} are realized.
This would provide a new class of signal for space-based gravitational wave observatories in the form of a stochastic ensemble of SMBH binaries, particularly for concepts
with longer detector arms and lower 
frequency sensitive bandwidths (see \eg \cite{Folkner}),
beyond the expected observation of individual sources sweeping through the band.

We note that the gravitational wave signal we have estimated does not truncate abruptly as might be assumed from Fig.~\ref{fig:h}.  Rather, the individual sources 
with $\Delta f > T^{-1}$ that we mentioned earlier will sweep through the band toward higher frequencies.
They will no longer constitute a stochastic signal, but will have a significant impact on the event rate of resolvable sources.  
Given the small fraction of total mergers represented in our Monte Carlo, we do not adequately represent these more rare
``chirping'' signals.  A much larger sample, or limiting the parameter range from which the sample is taken,
would allow us to better represent the full merger population and predict the event rate for individually resolvable merger events, 
which we plan for future work.  In addition to the expectation of an increased merger rate for SMBHs, the increased galaxy merger rate implies 
that a large number of satellite black holes that fail to merge with their host, and may be observable as ULXs.  This provides
a second definitive prediction for our model, and will be explored in the next section.

\section{Stalled black-hole mergers and ULXs}
\label{sec:ulx}
\begin{figure}
\includegraphics[trim = 0mm 0mm 0mm 0mm, width=0.48\textwidth]{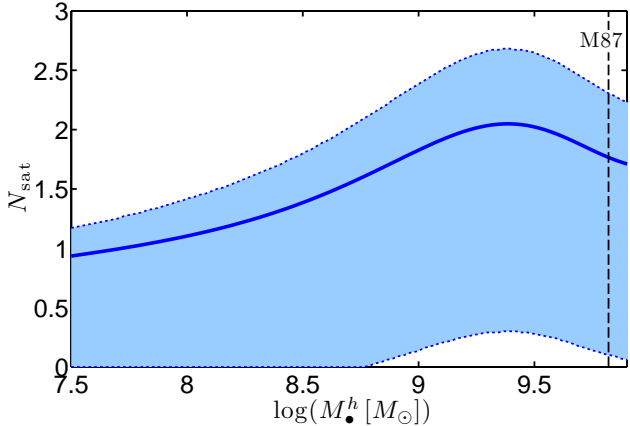}
\hfill
\caption{The expected number of satellite black holes, $N_{\rm sat}$, for a given host black-hole mass $M^h_{\bullet}$ is shown
with a solid (blue) line.  The central three quintiles of the probability distribution for $N_{\rm sat}$ are shaded and bound by
dotted (blue) lines.  The dashed (black) line shows the mass of M87 for reference.
\label{fig:nsat}
}
\end{figure}

We have noted that, for very massive galaxies such as BCGs, all but the most comparable mass galaxy mergers stall prior to black hole merger.
This has little effect on the
total gravitational wave signal, since it is dominated by major and near-major mergers.  
However, the possibility of ``stalled'' mergers, where a smaller satellite black hole resides in the outer regions of a massive host galaxy,
is an interesting candidate as a model for ULXs.  Indeed, several authors have studied the dynamics of stalled mergers and their viability as ULX sources 
\citep{Islam1,Islam2,Islam3,Islam4,VolPerna}.  Given the rarity of very massive 
galaxies, the probability of a merger with a comparable-mass galaxy is small.
Therefore, very massive galaxies will often retain some of the black holes from
merged satellite galaxies, since they will be unable to reach parsec-scale separations from the host black hole through dynamical friction in less than a Hubble time.  For an extremely massive system such as M87 with $M^h_{\bullet}=6.6\times 10^9\,\msun$, even very massive satellites
with $M^s_{\bullet} \lsim 5\times 10^9\,\msun$ would stall according to Eq.~\ref{eq:tdf}, so essentially any merger would result in a stalled satellite.
We can calculate the expected number of satellites for a given host mass, $\langle N_{\rm sat}(M^h_{\bullet})\rangle \equiv N_{\rm sat,\,total}(M^h_{\bullet})/N_{\rm host}(M^h_{\bullet})$, where
\begin{widetext}
\beq
N_{\rm sat,\,total}(M^h_{\bullet}) = \dint_{0}^{1}dz\, P(z)\dint_{\mlow}^{M^h_{\bullet}} \phi\,(M^h_{\bullet}) \,\phi_{\rm low}(M')\, \Theta \left[t_{\rm H} - t_{\rm DF} \left(M^h_{\bullet},M',z\right)\right]dM' \,,
\eeq
\end{widetext}
$\Theta(t)$ is 1 for $t>0$ and 0 otherwise, and 
\beq
N_{\rm host}(M^h_{\bullet}) = \dint_{0}^{1}dz\, \frac{dV_c/dz}{n(z)}\, \phi(M^h_{\bullet})\,.
\eeq
Since $N_{\rm sat}$ will be Poisson-distributed, calculating $\langle N_{\rm sat}(M^h_{\bullet})\rangle$ allows us
to calculate the full distribution of stalled satellites for a given host. 
We show the result in Fig.~\ref{fig:nsat}, including the expectation value and the middle
three quintiles of the distribution.  

It is interesting to note that
the mean number of satellites for a massive host galaxy is $\sim 1$--2, consistent with the observed population of
ULXs.  That these satellites would be found in the outer regions of the host galaxy, and would likely be accompanied
by enhanced star formation due to the interaction of the host and satellite bulges, is also consistent with the observed characteristics of ULXs.  It has been noted that multiple
ULXs are preferentially found in actively merging galaxies, so identified through morphology and kinematics (see \eg \cite{Kaaret} and references therein).
Because ULXs exceed the Eddington luminosity, $L_E$, for a $20\,\msun$ black hole, which is the largest mass expected through typical stellar evolutionary
channels \citep{Fryer}, intermediate-mass black holes have been proposed as ULX central engines \citep{Colbert:2002mi}.

In the noteworthy case of M82, which is likely to have undergone a recent merger, it has been suggested that the ULX M82 X-1
may be a low-luminosity active galactic nuclei powered by a intermediate-mass black hole (IMBH) from the core of a dwarf galaxy that M82 cannibalized
\citep{King}.  Alternatively, it has been suggested that ULXs are simply low mass X-ray binaries that are either accreting beyond the Eddington limit,
or whose emission is being beamed.
However, SMBH satellites, spiraling toward the host core and accreting at the Bondi-Hoyle-Lyttleton rate, would naturally explain
the luminosity of ULXs without resorting to super-Eddington accretion, or the as-yet-uncertain existence of IMBHs.  This possibility
has not been widely considered, given the common assumption that sufficiently massive satellites would sink to the host core
in less than a Hubble time.  However, as we have shown, this is far from the case in general, and especially so for BCGs.
The luminosity of a satellite black hole is given by
\begin{widetext}
\beq
L_B=\eta \dot{M}_B c^2 = 4\pi \eta \left(cGM^s_{\bullet}\right)^2 \rho_{\infty} v^{-3} 
=3\times 10^{39}\,{\rm erg/s}\,\left(\frac{\eta}{0.1}\right) \left(\frac{M^s_{\bullet}}{2.2\times 10^5\,\msun}\right)^2 \left(\frac{n}{100\,{\rm cm}^{-3}}\right) \left(\frac{v}{190\,{\rm km/s}}\right)^{-3}\,, 
\label{eq:bondi}
\eeq
\end{widetext}
where $\eta$ is the efficiency of converting accreted gas into luminosity, $\dot{M}_B$ is the Bondi-Hoyle-Lyttleton accretion rate, 
$\rho_{\infty}$ and $n$ are the background gas and particle number density, respectively, and $v\equiv \sqrt{v_{\bullet}^2 + \sigma^2 + c_s^2}$ 
is the total
relative velocity of the satellite and the background gas, with $v_{\bullet}$ the velocity of the satellite with respect to the mean
gas flow, $\sigma$ the local velocity dispersion, and $c_s$ the local sound speed.
We have normalized the mass so that $L_B$ matches the bolometric Eddington luminosity of a $20\,\msun$ black hole.

\begin{figure}
\includegraphics[trim = 0mm 0mm 0mm 0mm, width=0.48\textwidth]{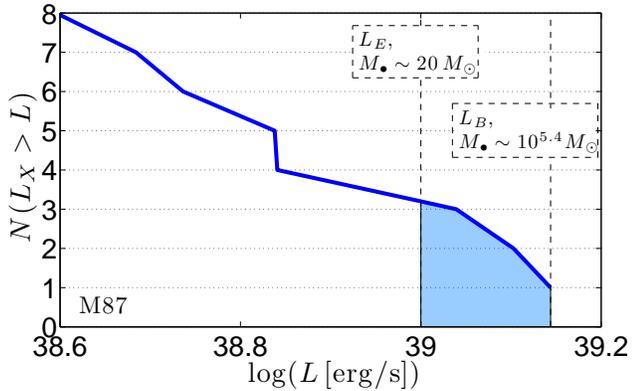}
\hfill
\caption{The luminosity distribution for M87 \citep{Jordan} is shown with a solid (blue) line.  The Eddington luminosity, $L_E$, of a 20 $\msun$ black hole,
and the Bondi-Hoyle-Lyttleton luminosity, $L_B$, of a $10^{5.4}\,\msun$ black hole are given by the dashed (black) lines.  The number of ultraluminous
X-ray sources is consistent with the number of SMBH satellites expected in M87, as shown in Fig.~\ref{fig:nsat}.
\label{fig:nlum}
}
\end{figure}

In Fig.~\ref{fig:nlum}, we show the tail of the luminosity function for M87, the most massive nearby example of a BCG.
We follow the convention of assuming that an X-ray luminosity of $10^{39}$ erg/s for a $20\,\msun$ black hole implies a total bolometric luminosity beyond
the Eddington limit.  We note that, under this assumption, M87 contains $\sim2$ ULXs, whose luminosity could be explained by the
presence of a SMBH satellite, although their luminosities are less extreme than M82 X-1, and could also be explained by an unusually
massive stellar-mass black hole or by marginally super-Eddington accretion.  
In the case of M82 X-1, the observed X-ray luminosity of $\sim 10^{41}$ erg/s
could be explained by the presence of a Sagittarius A*-sized satellite black hole spiraling through M82 under dynamical friction.
Any ULX could also potentially be explained by the aforementioned models of a super-Eddington or beamed emission
from a stellar-mass black hole, or an IMBH
accreting near the Eddington limit.  There is as-yet no unambiguous way to distinguish these scenarios from our supermassive
satellite scenario; for example, since the emission mechanism is unclear, we cannot link variability to mass as conclusively as is done
with active galactic nuclei.  Furthermore, although quasi-periodic oscillations are observed in some ULXs, they are not a clear indicator
of a binary companion, and could arise generically in black hole accretion (see \eg \cite{Dolence}). 
Our scenario has the advantage of corroborating observational evidence in the form of merging galaxies
and enhanced star formation in galaxies hosting ULXs, as well as requiring only conventional astrophysics in the form of
dynamical friction and Bondi-Hoyle-Lyttleton accretion to explain the observations.
The abundance of ULXs may therefore be an additional indicator of the merger-driven nature of the evolution of the mass function for $z\leq1$.

\section{Conclusions}
\label{sec:conc}
We have demonstrated that the mass function for massive galaxies since $z=1$ is consistent with merger-driven evolution.
Massive galaxies, though rarer than their smaller counterparts, are critical when considering overall luminosity, star formation rates,
gravitational wave emission, etc.~since the majority of total black hole mass is contained in
individual black holes with masses above $\sim 5\times 10^8\,\msun$ at $z=0$ (according to the mass function used in this work).  
Furthermore, despite the
domination of the mass function by the number of smaller galaxies,
our results indicate that the mass-weighted mean mass ratio between a host galaxy
and its satellite is $\sim 1/4$, which is consistent with \cite{Naab} and \cite{Oser}.
The novel expectation of at least one, and potentially several mergers with a comparable-mass companion for these massive galaxies
\citep{Trujillo,Hopkins} has dramatic implications
for gravitational wave observations.
We have shown that if this new paradigm of merger-driven evolution holds true, then 
the gravitational wave signal from the stochastic population of merging SMBH binaries is much
stronger than previously expected, exceeding previous signal-to-noise estimates by a factor of $\sim 2$--$5$, depending on the choose
of parameters in the initial mass function.  
Indeed, our mean estimate of $h_{\rm o} \geq 4.1\times 10^{-15}$ is near the 95\% confidence limit of
the European PTA ($6\times 10^{-15}$) and the NANOGrav PTA ($7\times 10^{-15}$), suggesting that PTAs may detect
gravitational waves in the very near future, with the caveat that a dip in the signal at the lowest observable frequencies, as evidenced
in many of our Monte Carlo realizations, would limit the sensitivity of PTAs, and also limit the rate at which the PTA constraint would
improve if analyzed optimally.

In addition to an expected increase in black hole mergers, our assumption of merger-driven galaxy evolution
also implies an increase in failed black hole mergers. 
For sufficiently massive host galaxies and/or
sufficiently disparate mass ratios between the host and satellite, dynamical friction will fail
to deliver the satellite to the core of the host.  These stalled satellites would be observed as luminous X-ray sources.
We find that, for the brightest cluster galaxy-component of the mass function, 1--2 of these stalled satellites
would be expected, with a large majority of hosts having 0--3 satellites.  We have shown that this result is 
consistent with existing observations of ultraluminous X-ray sources, such as those found in M87 and, notably, M82 X-1, although the nature of these sources is not yet known conclusively
and will require further study.  Nonetheless, a validation of our model through the observation of gravitational
waves with pulsar timing arrays in the immediate future would also strongly suggest that at least some of these
merging satellites would stall, and would therefore be observable as ultraluminous X-ray sources.

\acknowledgements
We thank Cole Miller, Priya Natarajan, and Alberto Sesana for helpful feedback on this manuscript.  STM thanks the Aspen Center for Physics for their hospitality during the final
stages of this work.

\bibliographystyle{apj}

\bibliography{/Users/sean/publish/bibtex/references}

\end{document}